\documentclass[aps,twocolumn,longbibliography]{revtex4-1}
\usepackage{amsmath,amssymb,latexsym}
\usepackage{graphicx}
\usepackage{bm}

\newcommand{\sd}{\downarrow}
\newcommand{\su}{\uparrow}

\begin{document}

\title{Ultrafast and reversible control of the exchange interaction in Mott insulators}
\author{J.~H.~Mentink}
\email{Johan.Mentink@mpsd.cfel.de}
\author{K.~Balzer}
\author{M.~Eckstein}

\affiliation{Max Planck Research Department for Structural Dynamics, University of Hamburg-CFEL, 22761 Hamburg, Germany}
\date{\today}

\begin{abstract}
The strongest interaction between microscopic spins in magnetic materials is the exchange interaction $J_\text{ex}$. Therefore, ultrafast control of $J_\text{ex}$ holds the promise to control spins on ultimately fast timescales. We demonstrate that time-periodic modulation of the electronic structure by electric fields can be used to reversibly control $J_\text{ex}$ on ultrafast timescales in extended antiferromagnetic Mott insulators. In the regime of weak driving strength, we find that $J_\text{ex}$ can be enhanced and reduced for frequencies below and above the Mott gap, respectively. Moreover, for strong driving strength, even the sign of $J_\text{ex}$ can be reversed and we show that this causes time reversal of the associated quantum spin dynamics. These results suggest wide applications, not only to control magnetism in condensed matter systems, for example, via the excitation of spin resonances, but also to assess fundamental questions concerning the reversibility of the quantum many-body dynamics in cold atom systems.
\end{abstract}

\maketitle

Controlling magnetically ordered systems on sub-picosecond timescales is currently a widely studied research area owing to the joint fundamental interest and technological demand for faster and more energy-efficient magnetic storage \cite{Kirilyuk10}. The fastest pathways to reverse magnetic order utilize the exchange interaction $J_\text{ex}$ between microscopic magnetic moments \cite{radu2011,mentink2012,wienholdt2013,baryakhtar2013,evans2014}, which can exceed external magnetic fields by orders of magnitude. Because $J_\text{ex}$ relies on the electrostatic Coulomb repulsion and the Pauli principle rather than on magnetic dipole forces, it may be modified directly by the action of a laser pulse on the electronic state. This implies an appealing and largely unexplored scenario to control magnetism on the fastest possible timescale. Recently, several experimental studies on magnetic materials have discussed an ultrafast modification of $J_\text{ex}$ or change of the type of magnetic order by creating a nonequilibrium electron distribution (by photo-doping or laser-heating)\cite{rhie2003,ju2004,thiele2004,wall2009,Forst11,carley2012,Li13,Matsubara13,mentink2014}. In these cases the spin dynamics after the excitation strongly depends on the relaxation of the electrons, thereby hindering a direct and reversible control of the spin degrees of freedom alone. On the other hand, reversible electrical control of $J_\text{ex}$ was recently demonstrated in a multiferroic solid state system, where the bond alignment can be changed by a static electric field \cite{ryan2013}. Clearly, a natural goal is to achieve control of $J_\text{ex}$ which is both reversible and ultrafast, \textit{i.e.}, it is active while a laser pulse is on, but leaves the electronic state unexcited after the pulse is switched off. 

A versatile approach to reversibly control the dynamics of quantum systems is given by the rectification of time-periodic perturbations. The use of periodic driving to control the dynamics of a quantum system is known in many areas of physics, \textit{e.g.}, through effective conservative forces resulting from the AC Stark effect, or through the coherent destruction of tunneling \cite{dunlap1986,grossmann1991}. For extended solid state systems, it is well known that particles in a tight-binding band subject to periodic driving evolve under an effective Hamiltonian that has a different band structure 
\cite{dunlap1986,grossmann1991,lignier2007,oka2009,tsuji2011,wang2013}. On the other hand, the control of the exchange interaction requires to understand how the driving influences both the band structure and the electronic correlations, which determine $J_\text{ex}$ in equilibrium. This is a highly challenging problem in general, since it implies the solution of a strongly time-dependent many-body problem of an extended system.

In recent years, important insights into the control of $J_\text{ex}$ have been obtained by studying the effect of periodic driving for one and two spin systems, leading to light-induced Kondo effects \cite{shahbazyan2000}, the design of an effective low-energy spin Hamiltonian in bosonic double-well systems\cite{chen2011}, as well as an optically induced RKKY interaction between localized spins in semiconductor quantum dots by virtual excitation of delocalized excitons \cite{piermarocchi2002,piermarocchi2004}. The latter has been shown to be effective in extended systems as well, in particular for the dilute ferromagnetic semiconductors \cite{fernandez-rossier2004,wang2007}. However, since the spin dynamics in ferromagnets requires a change of the total angular momentum, it is difficult to induce fast dynamics by modifying $J_\text{ex}$. To the contrary, extended antiferromagnetic systems do not suffer from the angular momentum bottleneck and hence can provide novel opportunities for the ultrafast control of the spin dynamics by modifying $J_\text{ex}$.

In this paper, we demonstrate that it is possible to reversibly control $J_\text{ex}$ in extended antiferromagnetic Mott insulators by periodically modulating the electronic structure with a frequency $\omega$ higher than $J_\text{ex}/\hbar$, but not resonant to electronic excitations. We investigate a simple driving scheme using time-periodic electric fields, which can be realized both in solid state systems and for cold atoms, and hence suggests wide applications: Besides the possibility of manipulating magnetism in solids, \textit{e.g.}, via the excitation of spin resonances, we find that in the extreme limit of strong driving one may even achieve a sign reversal of $J_\text{ex}$, which is equivalent of letting the system evolve backwards in time and may allow for addressing fundamental questions concerning the reversibility of quantum many-body dynamics \cite{lebowitz1993,polkovnikov2011,caneva2014} in cold atom experiments. Furthermore, we show that considerable insight can be obtained from analytical Floquet theory for a few-site cluster under continuous driving, which predicts reversible enhancement, reduction and even complete sign-change of the exchange interaction. The relevance of these results for extended many-body systems may not be clear \textit{a priori}, since in this case a true quasi-steady driven state may always become infinitely excited \cite{dalessio2014}. For the relatively short-term dynamics of interest here, the predictions of Floquet theory are nevertheless correct, as we demonstrate using numerical calculations for both the classical spin dynamics in high-dimensional Mott insulators and the quantum spin dynamics in low-dimensional Mott insulators.

\section{Results}

{\bf Floquet theory for a two-site cluster --}
In this work we study the repulsive Hubbard model as a model for strongly interacting electrons on a lattice. The Hamiltonian is given by
\begin{align}
\label{repulsive hubbard}
H
=&
-t_0
\sum_{\langle ij \rangle \sigma}
c_{i\sigma}^\dagger
c_{j\sigma}
+
U
\sum_{j}
n_{j\uparrow}n_{j\downarrow},
\end{align}
where $c_{i\sigma}^\dagger$ creates an electron at site $i$ with spin $\sigma=\su,\sd$, $t_0$ is the hopping between nearest-neighbor sites, and $U$ the repulsive on-site interaction. Arbitrary time-dependent electric fields $\mathbf{E}(t)$ are incorporated by adding a Peierls phase to the hopping matrix elements (see Methods). Below we set $\hbar=1$ and measure energy and time in units of the hopping $t_0$ and the inverse hopping, respectively. Electric fields are measured in units of $t_0/ea$, where $a$ is the lattice spacing and $e$ the electron charge.

\begin{figure}[t]
\includegraphics[width=\columnwidth]{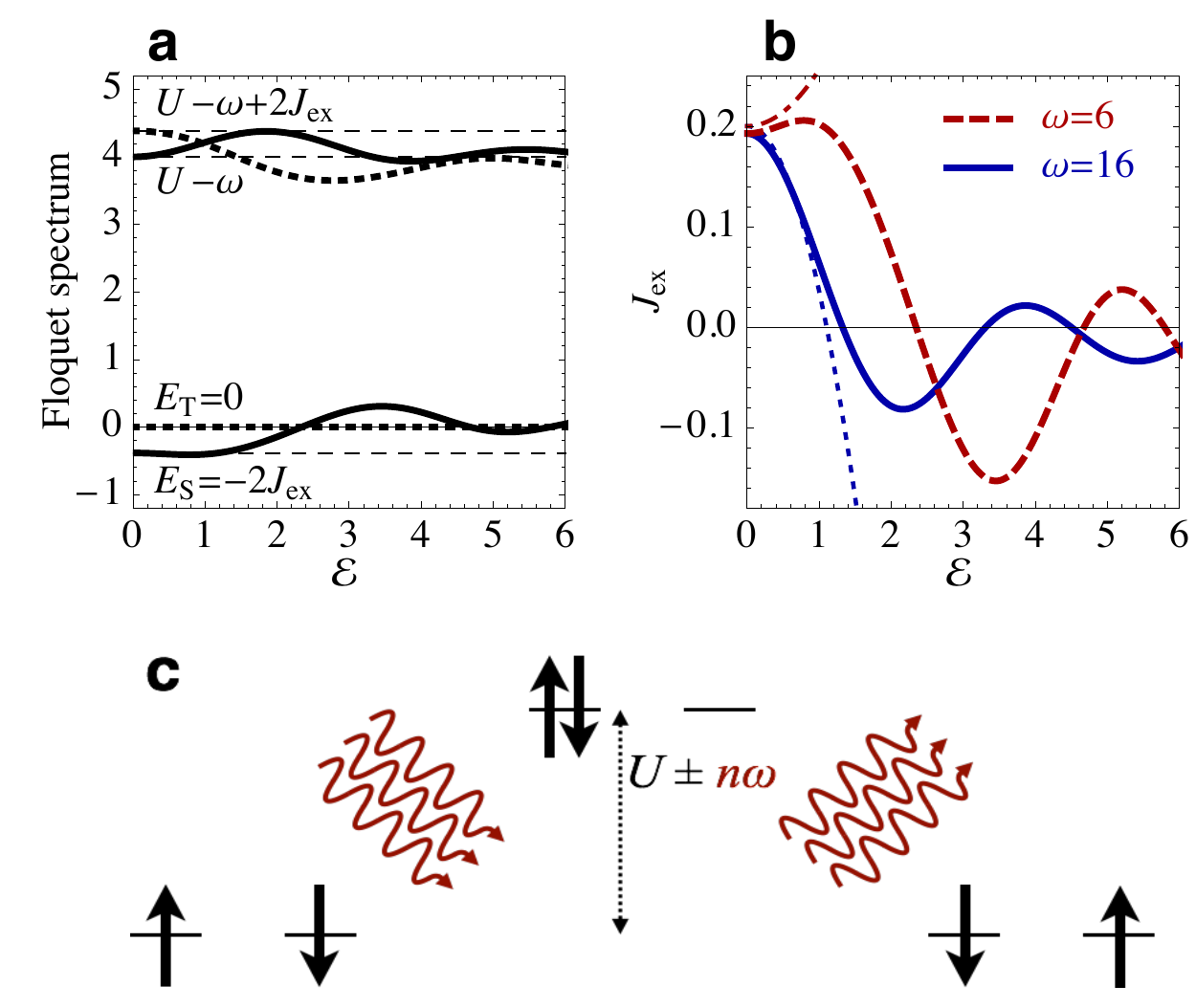}
\caption{
{\bf Floquet spectrum and extracted exchange interaction in a two-site Hubbard model}. 
({\bf a}) Floquet spectrum for $U=10$ and $\omega=6$ as a function of Floquet amplitude $\mathcal{E}$. Thin dashed lines indicate the spectrum without driving. The exchange interaction ($J_\text{ex}$) is extracted from the difference between the lowest singlet ($E_\text{S}$) and triplet ($E_\text{T}$) levels. The result is shown in ({\bf b}) for the same $U=10$ and two different frequencies above ($\omega=16$, blue solid line) and below ($\omega=6$, red dashed line) the Mott gap. For large driving strength $\mathcal{E}\gtrsim 1$, a reversal of $J_\text{ex}$ is possible. In addition, thin blue dotted ($\omega=16$) and thin red dash-dotted ($\omega=6$) lines indicate the modification of $J_\text{ex}$ as obtained within second order perturbation theory (Eq.~\eqref{pertb}). 
({\bf c}) Illustration of the modification of $J_\text{ex}$ in the Floquet picture. Red wavy lines indicate the virtual absorption and emission of an arbitrary number $n$ of photons 
with frequency $\omega$. This induces a coupling to excited states in different Floquet sectors displaced by an energy $U\pm n\omega$.
\label{fig1}}
\end{figure}

For half-filling and $U/t_0\gg1$, the Hubbard model describes a Mott insulator with one electron per site, in which the remaining spin degrees of freedom are coupled by an antiferromagnetic exchange interaction $J_\text{ex}=2t_0^2/U$. The simplest analytical understanding for this result is obtained already for two electrons on two Hubbard sites: For total $S_z=0$ we have four states. In the atomic limit two of them ($|\uparrow,\downarrow\rangle$ and $|\downarrow,\uparrow\rangle$) are singly occupied sites at $E_1=0$ while the other two states involve a doubly occupied and empty site at energy $E_2=U$ ($|\uparrow\downarrow,0\rangle$ and $|0,\uparrow\downarrow\rangle$). In the presence of hopping, the degeneracy is lifted and the lowest states become singlet and triplet states at energies $E_S=-4t_0^2/U$ and $E_T=0$, respectively. The low-energy spectrum is thus described by a spin Hamiltonian $2J_\text{ex} \mathbf{S}_1 \mathbf{S}_2$ with $J_\text{ex}=(E_T-E_S)/2 = 2t_0^2/U$. This analytical understanding from the cluster is useful since lattice effects beyond the lowest-order perturbative result only appear in the order $t^4_0/U^3$. In the same spirit, to gain theoretical insight into the modification of $J_\text{ex}$ by periodic driving, we first consider the same two-site Hubbard cluster and employ Floquet's theorem \cite{Floquet1883,GrifoniHanggi1998}, the analog of Bloch's theorem in time. When the Hamiltonian is periodic in time with a period $T=2\pi/\omega$, solutions of the time-dependent Schr\"odinger equation are given in the form  $| \psi (t) \rangle = e^{- \text{i}\epsilon_\alpha t } | \psi_\alpha (t) \rangle$ where $| \psi_\alpha (t+T) \rangle = | \psi_\alpha (t) \rangle$ is time-periodic, and  $\epsilon_\alpha$ is a quasi-energy defined up to multiples of $\omega$. The Floquet picture describes a system that undergoes virtual absorption and emission of an arbitrary number of photons, as depicted in Fig.~\ref{fig1}c for the Mott-Hubbard systems. The unperturbed Floquet sectors are described by the time-averaged Hamiltonian shifted by $n\omega$, and mixing between these Floquet sectors results in a renormalization of quasi-energy levels. A natural procedure is then to adopt an "adiabatic" principle in which the driving amplitude varies slowly as compared to the driving frequency, $|\dot{E}/E|\ll\omega$, and identify the amplitude-dependent singlet-triplet splitting $\epsilon_T-\epsilon_S$ in the quasi-energy spectrum with the (time-dependent) exchange interaction that describes the spin dynamics in the laser driven system on timescales much slower than the driving period $T$. 

The numerical solution of the Floquet spectrum and various analytically tractable limiting cases for the two-site Hubbard model are detailed in the methods section. For a tight-binding model driven by an electric field $E(t)=E_0\cos(\omega t)$, the coupling between Floquet sectors is controlled by the dimensionless Floquet amplitude
\begin{align}
\label{floquet amplitude}
\mathcal{E}= \frac{eaE_0}{\hbar \omega },
\end{align}
and the time-averaging of $H$ corresponds to a coherent reduction of the tunneling amplitude \cite{dunlap1986,grossmann1991} by a factor $J_0(\mathcal{E})$, where $J_0$ is the Bessel function (see Methods). In Fig.~\ref{fig1}, we display the Floquet spectrum and the exchange splitting $J_\text{ex}(\mathcal{E})$ for a half-filled two-site Hubbard  model. In contrast to the limit $\omega\gg U,t_0$, where the only effect would be a renormalization of the hopping by $J_0(\mathcal{E})$ and a corresponding reduction of the exchange splitting at large $U$ by a factor $J_0(\mathcal{E})^2$, one can see that $J_\text{ex}$ can be both increased and decreased for finite $\omega$, depending on the driving. This is clear already in the perturbative limit for $\mathcal{E}\ll1$ and $t_0/U\ll1$, which is given by 
$J_\text{ex}=2t_{0}^2/U + \Delta J_\text{ex}$ with (see Methods)
\begin{align}\label{pertb} 
\Delta J_\text{ex}
&=
\frac{\mathcal{E}^2
t_{0}^2}{2}
\Big(
\frac{1}{U+\omega}
+
\frac{1}{U-\omega}
-\frac{2}{U}
\Big),
\end{align}
and indicated with (dash-)dotted lines in the right panel of Fig.~\ref{fig1}. The last term of Eq.~\eqref{pertb} is the reduction of the exchange due to coherent reduction of the tunneling, while the first two terms derive from the coupling to the $m=\pm1$ Floquet sector with effectively shifted charge transfer energies $U\pm \omega$. The net effect is an enhancement (reduction) of $J_\text{ex}$ for driving frequencies below (above) the Coulomb energy $U$. For sufficiently strong driving one can even reverse the sign of $J_\text{ex}$, thus leading to the remarkable finding of a ferromagnetic exchange coupling in the half-filled Hubbard model. This happens when $\mathcal{E}$ is of order one, such that coupling to higher Floquet sectors with effective Coulomb energy $U-m\omega$ becomes strongly enhanced, while the direct exchange path is reduced by coherent destruction of tunneling $\sim J_{0}(\mathcal{E})$. For larger Floquet amplitudes, the direct exchange path again increases due to the oscillating behavior of the Bessel function. In the remainder of this paper we verify that these predictions from the two-site Floquet picture can remain valid for extended condensed matter systems at off-resonant driving with finite pulse duration, in spite of the possible importance of higher-order processes such as multi-photon absorption in the strongly driven regime, and the limited number of cycles in the pulse.

{\bf Mean-field spin dynamics in a high-D lattice --}
A direct prediction of the Floquet theory (Eq.~\eqref{pertb}) is the enhancement (reduction) of the exchange interaction for driving below (above) the Mott gap with weak amplitudes ($\mathcal{E}\ll1$). A large class of materials for which this might be relevant are three-dimensional bulk systems, such as transition metal oxides, whose low-energy spin dynamics contain uniform spin resonances that can be conveniently described in mean-field theory. For large dimensions, also a numerical solution of the nonequilibrium electron dynamics in the Hubbard model is possible using the dynamical mean-field theory (DMFT, see Methods). 

\begin{figure}[t]
\includegraphics[width=\columnwidth]{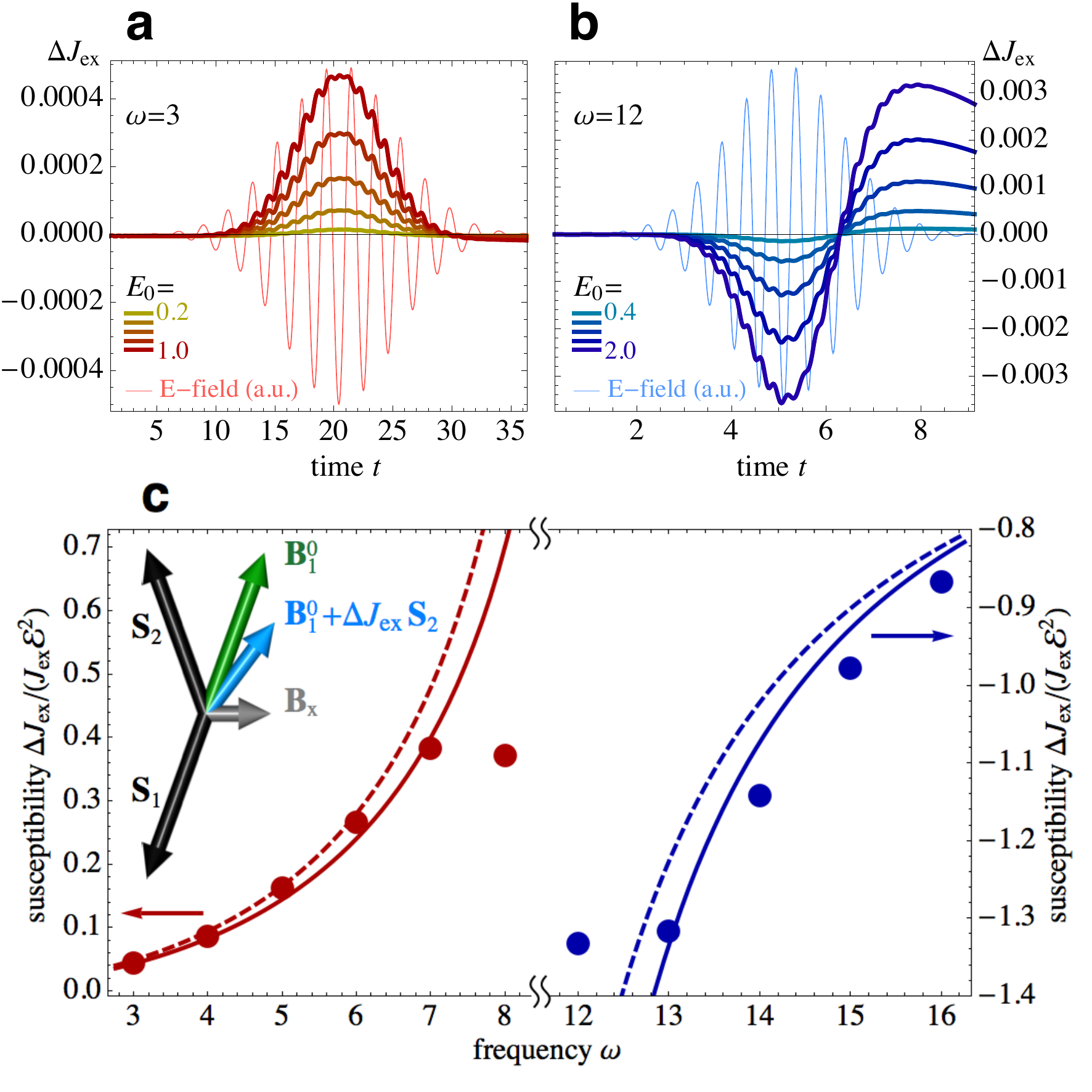}
\caption{
{\bf Laser-induced modification of the exchange in the Hubbard model on the hyper-cubic lattice}.
Panel {\bf a} and {\bf b}: Time-dependent change of the exchange interaction ($\Delta J_\text{ex}$, thick lines) during the action of a laser pulse, for driving frequencies $\omega=3$ ({\bf a}) and $\omega=12$ ({\bf b}) below and above the Mott gap, respectively. Different colours correspond to results obtained with different amplitude $E_0$ of the electric field,  increasing from light to dark. Thin lines show the time dependence of the electric field. Numerical results were obtained using Dynamical Mean Field Theory (DMFT) for the hyper-cubic lattice at $U=10$ and initial temperature $T=0.025$. {\bf c}) The driving susceptibility $\Delta J_\text{ex}/(J_\text{ex}\mathcal{E}^2)$ for $\mathcal{E}\to0$ for frequencies above (blue, right vertical axis) and below gap (red, left vertical axis), obtained from DMFT for the hyper-cubic lattice (disks), from the numerical Floquet spectrum of a two-site Hubbard cluster (solid lines), and from the perturbative result Eq.~\eqref{pertb} (dashed lines). The inset illustrates the canted geometry of the two sub-lattice magnetizations $\mathbf{S}_{1,2}$ (black arrows) induced by a static transverse magnetic field $\mathbf{B}_\text{x}$ (gray arrow). In equilibrium the effective magnetic field $\mathbf{B}_1^0$ (green arrow) is collinear with $\mathbf{S}_1$. A modification of the exchange interaction ($\Delta J_\text{ex}$) would rotate the effective field (light blue arrow) with respect to $\mathbf{S}_1$ causing the excitation of a spin resonance. In the DMFT calculations, $\Delta J_\text{ex}$ is computed from the observed spin precession in this canted geometry (see Methods). 
\label{fig2}}
\end{figure}

Within DMFT, the equilibrium solution of the Hubbard model at half-filling and low temperature is the N\'eel state. In order to asses the exchange interaction in this state, we study the excitation of resonances in the antiferromagnetic phase in a transverse magnetic field $B_x$,
a setup that was pioneered in \cite{mentink2014} for resonant photo-excitation. In equilibrium, the balance of $B_x$ and $J_\text{ex}$ gives rise to a canting of the  magnetic sublattices out of the $y$-$z$ plane. If $J_\text{ex}$ is modified under periodic driving, the sublattice magnetizations are no longer aligned with the effective field $B_\text{eff}$ given by external and exchange fields, as illustrated in the inset of Fig.~\ref{fig2}. This implies a rotation of the spins in the plane perpendicular to $B_x$ (leaving the total angular momentum $S_x$ conserved), from which the time-dependent modification of the exchange interaction is calculated (see Methods). Hence, in our calculations $J_\text{ex}$ is defined as the parameter that describes best the observed transverse spin dynamics as is obtained from the solution of the full electron problem. We stress that this allows us to quantify changes of the exchange interaction independent of the exchange energy stored in the system. This is particularly important for the regime in which absorption is not negligible, where the laser excites mobile carriers which transfer their energy to the spin background on an ultrafast timescale and thus reduce the ordered moment. 

We have implemented the DMFT solution of the Hubbard model in a time-dependent external electric field for the infinite-dimensional hyper-cubic lattice with density of states $D(\epsilon)=\exp(-\epsilon^2)/\sqrt{\pi}$ \cite{aoki2014rmp,Turkowski2005a}. The electric field is pointing along the body diagonal of the lattice and represents a laser pulse with frequency $\omega$ and a Gaussian envelope that contains 15 cycles per pulse, \textit{i.e.}, $E(t)=E_0\sin(\omega t)\exp(-(t-3t_c)^2/t_c^2)$ with $t_c=15\pi/(2.1\omega)$. Fig.~\ref{fig2}a and b show time traces of the induced change of the exchange interaction $\Delta J^c_\text{ex}(t)$ for one driving frequency below (a) and above (b) gap, as extracted from the time evolution of the spin degrees of freedom during the pulse. In accordance with the prediction of the Floquet theory we observe an enhancement (reduction) of the exchange interaction during the application of the field with a frequency below (above) gap. The frequency $\omega=3$ in Fig.~\ref{fig2}a is far from the resonance $\omega\approx U$, and we observe that $\Delta J^c_\text{ex}\approx 0$ after the pulse, demonstrating the reversibility of the effect. Conversely, the driving frequency $\omega=12$  in Fig.~\ref{fig2}b is chosen close to the edge of the upper Hubbard band where we observe significant absorption and transient behavior after the pulse. Hence, although the exchange interaction is modified in this case as well, the effect is not reversible. Note also that the time reached in the present simulations is too short for the photo-excited carriers to relax, hence a description in terms of a quasi-stationary photo-doped state discussed earlier \cite{mentink2014} is not yet valid. 

A quantitative comparison with the Floquet theory is shown in the bottom panel of Fig.~\ref{fig2}, where the ``driving susceptibility'' $\Delta J_\text{ex}/(J_\text{ex}\mathcal{E}^2)$ for $\mathcal{E}\to0$ is plotted as a function of the driving frequency. Solid discs show the DMFT results as obtained by a linear fit through the dependence of the ratio $\Delta J_\text{ex}(t)/J_\text{ex}$ on $\mathcal{E}^2$ at its maximum. Dashed and solid lines show the results based on the perturbative Floquet formula (Eq.~\eqref{pertb}) and the full Floquet spectrum (non-perturbative in $t_0/U$), evaluated from the derivative $dJ_\text{ex}/d\mathcal{E}^2$ at $\mathcal{E}=0$. As expected, in the vicinity of the band edge ($|\omega-U|\sim 2$), strong deviation is found since band absorption is not captured in a cluster picture. Away from the band edge, however, the frequency dependence matches very well, being even in quantitative agreement for the lowest frequencies below gap. This demonstrates the usefulness of the Floquet theory for understanding how off-resonant periodic driving modifies the exchange interaction in extended condensed matter systems by photo-assisted hopping. 

{\bf 1D quantum spin dynamics --} An intriguing prediction of the Floquet analysis is the existence of amplitude and frequency ranges in which the exchange coupling becomes ferromagnetic (FM). Such a sign change of $J_\text{ex}$ cannot cause a transition to a FM  state since the Hubbard model Eq.~\eqref{repulsive hubbard} conserves the total spin. However, even if the system remains antiferromagnetic (AFM), a change of sign of $J_\text{ex}$ by periodic driving allows for a very non-trivial and unique way to control the spin dynamics, namely, to reverse the time evolution of the undriven system. Such time reversal can be anticipated by considering a pure Heisenberg spin Hamiltonian $H_\text{ex}=J_\text{ex}\sum_{\langle ij\rangle}{\mathbf{S}}_i{\mathbf{S}}_j$, which gives an accurate description of the low-energy spin dynamics in the half-filled Hubbard model at $U\gg t_0$ if the system is not electronically excited. In the absence of driving, the propagation over a time interval $t$ is given by the evolution operator $\mathcal{U}_{\text{AFM}}=\exp(-\text{i}H_\text{ex}t)$. Such evolution can exactly be reversed by the propagation with an exchange interaction $J_\text{ex}^\prime$ of opposite sign over a time interval $t^\prime=|J_\text{ex}/J_\text{ex}^\prime|t$, since for the FM time evolution operator we have $\mathcal{U}_{\text{FM}}=\exp(-\text{i}H_\text{ex}^\prime t^\prime)=\exp(+\text{i}H_\text{ex}t)=\mathcal{U}_{\text{AFM}}^{-1}$, \textit{i.e.}, the two time evolution operators are exactly inverse to each other.

\begin{figure}[tbp]
\includegraphics[width=\columnwidth]{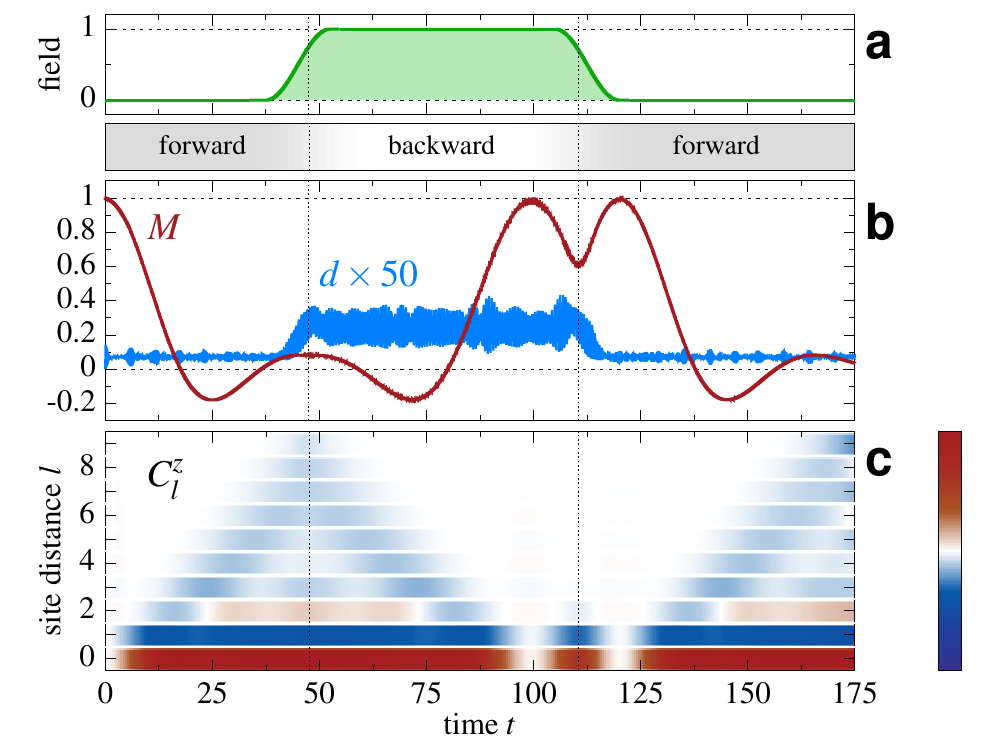}
\caption{
{\bf Time-reversal of the spin dynamics by periodic driving of a 10-site Hubbard chain}. {\bf (a)} Field envelope with cosine-shaped ramps of length $\Delta t=15$ around $t_\text{f}=45$ and $t=112.5$. The bar below the field envelope indicates forward (gray) and backward (white) time evolution when the field is off and on, respectively. {\bf (b)} Time evolution of staggered magnetization $M$ (red) and total double occupation $d$ (blue), scaled by a factor of $50$, showing free evolution for times $t<37.5$ and $t>120$ and evolution under an additional periodic driving at frequency $\omega/U=0.6$ and Floquet amplitude $\mathcal{E}=3.4$ in between. {\bf (c)} Build up and diminishing of the spin-spin correlation function $C_l^z$. Numerical results were obtained by exact diagonalization for $U=50$ and open boundary conditions, starting from a classical N\'eel state. The colour bar indicates the value of $C_l^z$, ranging from $-0.25$ (dark red) through $0$ (white) to $0.25$ (dark blue).
\label{fig3}}
\end{figure}

To demonstrate that periodic driving of the Hubbard model at large $U$ indeed yields the anticipated time reversal of the spin degrees of freedom, we consider a chain of $L=10$ sites and compute the dynamics using exact diagonalization techniques (see Methods). The system is initially prepared in a classical N\'eel state  $c_{1\uparrow}^\dagger c_{2\downarrow}^\dagger c_{3\uparrow}^\dagger \ldots |0\rangle$ and is evolved under the unperturbed Hamiltonian~\eqref{repulsive hubbard}. In a quantum Heisenberg model, the classical N\'eel state is a highly excited state the energy of which exceeds the thermal energy at the N\'eel temperature, such that no remanent long-range order is expected at long times, apart from finite size effects. In one dimension, not even the ground state displays long-range order. As a consequence of the spin-flip terms $J_\text{ex}(S_{i}^+S_{i+1}^-+S_{i}^-S_{i+1}^+)$ in the effective antiferromagnetic Heisenberg model, we thus observe a rapid decay of the total staggered magnetization $M=\frac{1}{L}\sum_{i=1}^{L}(-1)^{i+1}\langle n_{i\uparrow}-n_{i\downarrow}\rangle$ (Fig.~\ref{fig3}b). After this initial free evolution to a state where long-range order is suppressed, we ramp on a time-periodic electric field (Fig.~\ref{fig3}a), with Floquet amplitude $\mathcal{E}=3.4$ and frequency $\omega/U=0.6$ such that the Floquet theory for a two-site model predicts a reversal of the exchange coupling. Under the periodic driving one indeed observes a near perfect reversal of the dynamics of $M(t)$ in Fig.~\ref{fig3}b, which almost completely recovers to the initial value $M(t=0)$ around $t\approx 100$. Subsequently, $M(t)$ is reduced again by further evolution in the reverse direction, as a consequence of the spin-flip terms in the ferromagnetic model. This continues until the field is ramped off, after which one observes that the free evolution brings the system again back to the initial state, from which the same rapid decay of $M(t)$ is observed as for the initial free evolution. Hence, we conclude that the periodic driving allows for a reversible control of the spin dynamics for the timescale considered in our simulations. This is further confirmed by the time evolution of the total double occupation $d=\sum_{i=1}^L\langle n_{i\uparrow}n_{i\downarrow}\rangle$, which has the same mean value before and after the driving, demonstrating that electronic excitations due to the driving are negligible. The weak oscillations in $d(t)$ are caused by switching on the hopping at $t=0$, while the increased mean value of $d(t)$ during driving is due to photo-assisted hopping processes.

Time reversal can be demonstrated not only on the level of local observables. Figure~\ref{fig3}c displays the evolution of the spin-spin correlation function $C^z_l=\sum_{|i-j|=l}\frac{1}{N_l}(\langle S^z_{i} S^z_{j} \rangle-\langle S^z_{i}\rangle\langle S^z_{j}\rangle)$ as a function of distance $l$ and time ($N_l$ is the number of site pairs with distance $l$). Starting from the initial uncorrelated product state, correlations build up under the evolution of $H$. Even though the system is small, the spreading of correlations resembles the light-cone effect which has been observed in quantum many-body systems after a quench \cite{cheneau2012}, \textit{i.e.}, correlations stay zero outside the light cone $|l|\le2vt$, where $v$ is a maximal mode velocity \cite{Calabrese2006}, while short-range antiferromagnetic correlations emerge inside the light cone. Further, under the action of the periodic driving, the spin-spin correlations diminish with the same speed, restoring the initially uncorrelated state. 

\begin{figure}[t]
\includegraphics[width=\columnwidth]{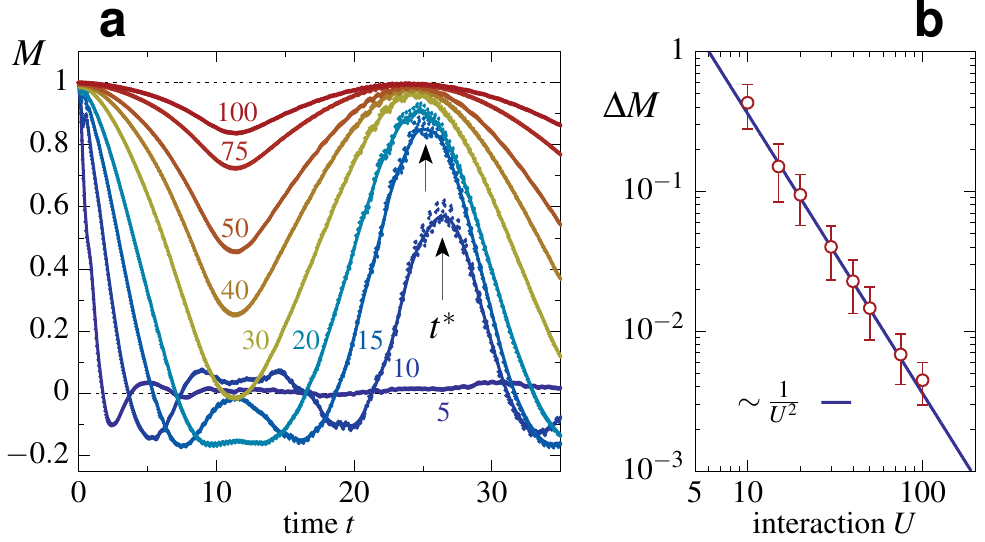}
\caption{ 
{\bf Quality of the time reversal}. {\bf (a)} Time-evolution of the staggered magnetization $M$ in a 10-site Hubbard chain, where after fixed forward propagation time $t_\text{f}=10$ the field is ramped up within a time interval $\Delta t=10$. Different colours correspond to different values of $U$, as indicated by the corresponding coloured labels. {\bf (b)} Difference $\Delta M$ between the initial staggered magnetization and the staggered magnetization at the revival time $t^*$ as a function of $U$. The error bars represent the magnitude of the the short-time fluctuations of $M$ close to $t^*$. In all calculations the driving frequency $\omega/U=0.6$ and Floquet amplitude $\mathcal{E}=3.4$ was used.
\label{fig4}}
\end{figure}

To determine quantitatively how well the time evolution is reversed in our simulations, we computed the difference $\Delta M = 1-M(t^*)$ between the initial magnetization in the N\'eel state and the magnetization $M(t^*)$ at the revival time $t^*$ for different values of $U$ (Fig.~\ref{fig4}). In all simulations the system is evolved forward in time for a given time $t_\text{f}=10$, after which the field is ramped on for a period $\Delta t=10$. As before we choose $\mathcal{E}=3.4$ and $\omega=0.6U$, which gives the same relative change of $J_\text{ex}$ for all sufficiently large values $U\gg t_0$. The observed scaling $\Delta M\sim1/U^2$ indicates that in the current setup the deviation from perfect reversal originates from small electronic excitations above the gap, which arises from switching on the hopping in the beginning of the simulation and the ramping on of the field. Since the dynamics of the electronically excited states is not captured by the spin Hamiltonian, it is not time-reversed. While the electronic excitation can be further reduced by slow ramping, the ultimate limit is given by non-Heisenbergian terms in the effective spin Hamiltonian, for which perfect reversal under periodic driving is not expected. For the half-filled Hubbard model in equilibrium, the leading-order correction to the Heisenberg model appears in the order $t_0^4/U^3$ in the strong-coupling expansion. At least for small times $t$, the contribution of a perturbation $\delta H$ proportional to $t_0^4/U^3$ to the time-reversed Hamiltonian would lead to a scaling $\Delta M \sim 1/U^3$, which is smaller than the electronic excitation in the present case. Note, however, that the times reached in the current simulations are nevertheless long enough to observe near perfect reversal even from a state without magnetic order, in which spin correlations have spread throughout the full chain (Fig.~\ref{fig3}). 

{\bf Discussion --} Our results demonstrate ultrafast and reversible electrical control of the exchange interaction in extended fermionic many-body systems by modulation with time-periodic electric fields. We emphasize that Floquet amplitudes $\mathcal{E} \sim 0.1$ are well accessible for condensed matter systems, which would lead to relative changes of $J_\text{ex}$ up to $1\%$. For example, for a frequency $\hbar\omega=1$eV and a lattice spacing of $2$\AA, a laser fluence of $1$mJ/cm$^2$ in a $100$fs pulse corresponds to a Floquet parameter $\mathcal{E}=0.05$. While realistic condensed matter systems usually involve several correlated bands, we think that our current results are already quite robust for the antiferromagnetic oxides governed by superexchange interactions. Similar as in the single band model studied here, superexchange interactions are governed by virtual charge excitations, which will be reversibly modified through the mechanism of photo-assisted hopping between different Floquet sectors. This is further supported by recently presented experiments on canted antiferromagnetic oxides \cite{mikhaylovskiy2014arx} using THz emission spectroscopy \cite{mikhaylovskiy2014}, which show the first experimental evidence of reversibly controlling exchange interactions by off-resonant pumping below the charge-transfer gap. Furthermore, Eq.~\eqref{pertb} implies a strong enhancement of off-resonant effects at low frequencies. As an extreme limit of this one can anticipate control of spin dynamics by few-cycle THz pulses, using both the coupling between the spins to the magnetic field of the light \cite{kampfrath2011}, and the modifications of $J_\text{ex}$ predicted by our work. At the same time, extensions to multi-band models are very interesting and relevant to perform as they usually involve multiple possibly competing exchange paths, which generally yield quantitative differences as well as different dependencies on the driving frequency due to the presence of additional resonances, potentially enabling to achieve even stronger effects. In this connection we also mention that extensions to multiband models may be relevant too for the description of metallic ferromagnets. Here static electronic structure calculations already indicate that exchange integrals involving electrons excited to higher bands considerably differ from those in the ground state \cite{zhang2014}. Furthermore, a more accurate quantitative description of photo-excited states would also include the effect of dynamic screening. However, we anticipate that this will be important only if mobile carriers are injected, i.e., for photo-doping excitations, while it will be a secondary effect for the off-resonant driving investigated here, which leaves the electron distribution unchanged.

While the single band model is clearly a minimal model for application to condensed matter systems, fermionic cold-atom systems resemble the single band model very accurately. Moreover, such systems may realize the large amplitudes $\mathcal{E}$ needed to achieve the ferromagnetic exchange at frequencies sufficiently far from resonances $n\omega=U$ which are difficult to realize for most condensed matter systems. It will therefore be of fundamental interest to investigate the reversal of the exchange interaction and the associated time reversal of the quantum spin dynamics in cold atom experiments. In these systems recently great progress has been made to prepare and measure systems with single-site spatial resolution \cite{bakr2010,endres2013}. Furthermore, various single-particle Floquet Hamiltonians could be realized in the limit of strong periodic driving without substantial heating by inter-band absorption (see, e.g. \cite{struck2012}). Cold atoms in optical lattices have been successfully used as a quantum simulator for the dynamics of a quantum quenches in the Bose-Hubbard model, starting from an artificially prepared charge-ordered phase \cite{Trotzky2012}, which suggest similar techniques to probe the behavior of spin systems under time reversal. In the methods section (cf.~Eq.~\eqref{modulation hopping}), we show that analogous time reversal can be achieved by modulating the amplitude of the hopping instead of its phase, which is easier to control in cold atoms. An intriguing problem to study both theoretically and experimentally is the fundamental question how well the time evolution can be reversed after (much) longer forward evolution time and investigate systematically the influence of small deviations from perfect time-reversal. Furthermore, a study of the Loschmidt echo and dynamical phase transitions \cite{heyl2013} in cold atoms might be possible by including additional perturbations to the back propagation. 


\section{METHODS}
\subsection{Floquet theory}

The Floquet quasi-energy spectrum can be obtained from the ansatz $| \psi (t) \rangle = e^{-\text{i}\epsilon_\alpha t } | \psi_\alpha (t) \rangle$ by expanding  $| \psi_\alpha (t) \rangle$ in a Fourier series  $| \psi_\alpha (t) \rangle=\sum_{m} e^{\text{i}\omega m t} | \psi_{\alpha,m} \rangle$, where $| \psi_{\alpha,m} \rangle$ is referred to as the component of the wave function in the $m$-th Floquet sector. The Schr\"odinger equation then achieves a block-matrix structure 
\begin{align}
\label{floquet matrix}
(\epsilon_\alpha + m\omega) | \psi_{\alpha,m} \rangle = \sum_{m'} H_{m-m'} | \psi_{\alpha,m'} \rangle,
\end{align}
where $H_{m} = (1/T) \int_0^{T} dt \, e^{\text{i}\omega m} H(t)$ are the Fourier components of the Hamiltonian. Different from the usual discussions of Floquet theory for single-particle Hamiltonians, here we focus on the effect of periodic driving on an electronic spectrum that is controlled by electronic correlations.

Time-dependent electric fields are incorporated into the Hubbard Hamiltonian \eqref{repulsive hubbard} by adding a time-dependent Peierls phase to the hopping matrix elements, $t_{ij}(t) = t_0 \exp[\text{i}{ea}A_{ij}(t) ]$, where $A_{ij}$ is the projection of the vector potential along the direction from site $i$ to $j$ (choosing a gauge with zero scalar potential and $E(t) = -\partial_t A(t)$). For the one-dimensional chain with electric field $E_0 \sin(\omega t)$ along the chain, this implies  $A_{ij}(t) = -\frac{1}{\omega} (i-j) E_0 \cos(\omega t)$. The Fourier components of the Hamiltonian are thus given by
\begin{align}
\label{perturbation3siteFT}
H_{m}
&=
-t_0
\sum_{\langle ij \rangle \sigma}
(-1)^m
J_m\big((i-j)\mathcal{E}\big)
c_{i\sigma}^\dagger
c_{j\sigma},
\end{align}
plus the additional (time-independent) interaction part in the $m=0$ component, where $J_m(x)$  is the $m$-th Bessel function, and the dimensionless parameter $\mathcal{E}= eaE_0 /(\hbar \omega )$ measures the strength of the perturbation [cf.~Eq.~\eqref{floquet amplitude}]. For the numerical determination of the Floquet energies, one truncates the number of Floquet sectors in Eq.~\eqref{floquet matrix} to $|n| \le N$, and increases $N$ to reach convergence. The determination of a many-body Floquet spectrum  thus requires the diagonalization of a matrix of dimension $N\times D$ where $D$ is the dimension of the Hilbert space. The results presented in Fig.~\ref{fig1} are converged with $N=8$.

In the limit of large frequency, $\omega \gg U,t_0$, Floquet sectors in Eq.~\eqref{floquet matrix} are separated in energy, and one can restrict oneself to the lowest sector $m=0$. This is equivalent to replacing the Hamiltonian with its time average, which leads to the renormalization of the hopping by $J_0(\mathcal{E})$, and a corresponding reduction  of the exchange by a factor $J_0(\mathcal{E})^2$. In the perturbative limit where both $t_0/U\ll 1 $ and $\mathcal{E} \ll 1$, we expand the Bessel functions $J_{n}(x)\sim x^n$ for $x\to 0$. To lowest order only states of the $m=0$ and $m=\pm 1$ Floquet sectors have to be taken into account, and the result given by Eq.~\eqref{pertb} follows from standard second order perturbation theory. Furthermore, an interesting limit for the Mott regime is given by $t_0/U\ll 1$, but allowing for fields of arbitrary amplitude. Because all terms $H_m$ for $m\neq 0$ are proportional to $t_0$, the perturbative shift of the spin states in the $m=0$ sector is given by a sum over all second-order processes containing precisely one virtual hopping to a higher Floquet sector and back, yielding
\begin{align}\label{dJexlargeU}
\frac{J_\text{ex}(\mathcal{E},\omega) }{J_\text{ex}(\mathcal{E}=0)}
= 
\sum_{n=-\infty}^\infty
\frac{J_{|n|}(\mathcal{E})^2}{1+n\omega/U}.
\end{align}
The unperturbed exchange is modified by a factor dependent only on $\omega/U$ and $\mathcal{E}$. For the parameters $\mathcal{E}=3.4$ and $\omega/U=0.6$ chosen for the time reversal, \textit{e.g.}, the factor is given by $-0.95$ and indicates a near perfect sign reversal. 

Finally, we note that a similar analysis is possible for the case of periodic modulation of the hopping amplitude, taking $t_0(t) = t_0(1+A\cos(\omega t))$. As above, we obtain for $U\ll t$ 
\begin{align}\label{modulation hopping}
\frac{J_\text{ex}(A,\omega) }{ J_\text{ex}(A=0) }
=
1+\sum_{n=\pm1} \frac{A^2}{1+n\omega/U},
\end{align}
yielding, \textit{e.g.}, a perfect sign reversal for driving above gap $\omega/U=1.2$ at $A\approx 0.67$. Other than 
for the field driven case, the perturbation is purely harmonic, and only Floquet sectors $m=\pm 1$ enter this expression.

\subsection{Dynamical mean-field theory}
To solve the electron dynamics in the Hubbard model we use nonequilibrium dynamical mean field theory (DMFT) \cite{Freericks2006,aoki2014rmp}. The electric field of the laser is incorporated by the Peierls substitution (see Method part A), so that the light matter interaction within the single band model is treated to all orders. Within DMFT \cite{Georges96}, which becomes exact in the limit of infinite dimensions \cite{Metzner1989}, local correlation functions are obtained from an effective impurity model in which one site of the lattice is coupled to a non-interacting, self-consistently determined bath. The impurity model is solved within the perturbative hybridization expansion (non-crossing approximation, NCA \cite{Eckstein2010nca}). The accuracy of this approach has been tested in equilibrium and for the short-time dynamics by comparison with higher-order hybridization expansions as well as with numerically exact Quantum Monte Carlo \cite{Eckstein2010nca, Eckstein2012c, Werner2012afm}, which revealed good agreement both in the antiferromagnetic and paramagnetic Mott insulator regime $t_0\ll U$. A detailed description of the formalism and of our numerical implementation is given in Ref.~\cite{mentink2014}, which studies the same setup (\textit{i.e.}, a hypercubic lattice with electric field along the body diagonal), yet for different electric fields.

\subsection{Determination of $J_\text{ex}$ in DMFT}
In order to asses the exchange interaction from mean-field spin dynamics, we investigate the antiferromagnetic phase of the Hubbard model supplemented with a term $B_x\sum_iS_{ix}$, which couples the spin $S_{i\alpha}=\frac{1}{2} \sum_{\sigma\sigma'} c_{i\sigma}^\dagger (\hat \sigma_\alpha)_{\sigma\sigma'} c_{i\sigma'}$ to a homogeneous magnetic field $B_x$ along the $x$-axis ($\hat \sigma_\alpha$ denote the Pauli matrices; $\alpha=x,y,z$). DMFT allows us to compute the time-dependent expectation value of the electron spin $\langle \mathbf{S}_{1,2}\rangle$ on the two magnetic sublattices. Assuming a rigid macrospin model, the time-dependent exchange interaction can be inferred from these results by inverting the Landau-Lifshitz equation  for the dynamics of spins on the mean field ${\mathbf{B}}_\text{eff}^{1,2} = B_{x}{\mathbf{e}}_x - 2 J_\text{ex} \langle \mathbf{S}_{2,1}\rangle$. It was shown \cite{mentink2014} that this approach compares well to the definition of exchange interactions from a time-dependent response formalism \cite{Secchi13}, as well as to the analytical perturbative result $J_\text{ex}=2t_0^2 /U$ in equilibrium at large $U$.  In the transverse field, the equilibrium exchange interaction can be determined from the canting induced by $B_x$, yielding $J_\text{ex}^c=-B_x/(4\langle S_x\rangle)$. Out of equilibrium, we obtain $J^c_\text{ex}(t)=-B_x/(4\langle S_x\rangle) +  \Delta J_\text{ex}^c(t)$,
\begin{equation}
\Delta J_\text{ex}^c(t) = -\frac{1}{4T\langle S_{1x}\rangle}\int_{t-T/2}^{t+T/2}\frac{\langle \dot{S}_{1y}(s)\rangle}{\langle S_{1z}(s)\rangle }ds,
\end{equation}
where the time-averaging is done to extract only the low-frequency component, similar as in the Floquet theory. Note that by calculating the exchange interaction in this way, we have automatically projected out dynamical changes in the (time-averaged) local magnetization $|\langle\mathbf{S}_{1,2}\rangle|$, and slight changes of the local moments (double occupation) as a result of the virtual absorption and emission of photons.

\subsection{Exact Diagonalization}
To compute the time evolution of the (driven) one-dimensional Hubbard model from the Schr\"odinger equation with a time-dependent Hamiltonian  $H(t)$ and a given initial state $|\psi_0\rangle=c_{1\uparrow}^\dagger c_{2\downarrow}^\dagger c_{3\uparrow}^\dagger \ldots |0\rangle$ we use the Krylov technique~\cite{hochbruck1997} in combination with a commutator-free exponential time-propagation (CFET) scheme~\cite{alvermann2011}. While the Krylov method provides efficient approximations to the time-propagator, which are important to treat large Hilbert spaces, the CFET scheme is related to the Magnus expansion and, preserving unitarity, allows for a high-order accurate integration of the Schr\"odinger equation in time. 

{\bf Acknowledgments --}
We thank U. Bovensiepen, M.I. Katsnelson, A.V. Kimel, A. Lichtenstein, R.V. Mikhaylovskiy, T. Oka, A. Secchi, N. Tsuji, and Ph. Werner for fruitful discussions. The exact diagonalization calculations were run in part on the supercomputer HLRN-II of the North-German Supercomputing Alliance. J.H.M. acknowledges funding from the Nederlandse Organisatie voor Wetenschappelijk onderzoek (NWO) by a Rubicon Grant.




%

\end{document}